\documentstyle[12pt]{article}
\newcommand{\OI}{{\cal O}_I}
\newcommand{\EI}{{\cal E}_I}
\newcommand{\OZ}{{\cal O}_0}

\begin{document}
\title{ Effective Hamiltonians with Relativistic Corrections\\
{\large\it I)\enskip
The Foldy--Wouthuysen transformation versus
the direct Pauli reduction}}
\author{H.~W.~Fearing, G.~I.~Poulis and S.~Scherer\\
TRIUMF, Vancouver, British Columbia, Canada V6T 2A3}
\date{}
\maketitle
\begin{abstract}
Two different methods of obtaining  ``effective $2\times 2$ Hamiltonians''
which include relativistic corrections to nonrelativistic calculations
are discussed.
The standard Foldy--Wouthuysen transformation generates Hamiltonians
which order by order in $1/M$ decouple the upper from the lower components.
The upper left--hand block then defines an effective $2\times 2$
Foldy--Wouthuysen Hamiltonian.
In the second method the matrix element of the interaction Hamiltonian of the
Dirac representation is evaluated between free positive--energy states
and reduced to two--component form.
The resulting expression (possibly expanded in $1/M$) then defines what we call
the ``direct Pauli reduction'' effective $2\times 2$ Hamiltonian.
We wish to investigate under which circumstances the two approaches yield the
same result.
Using a generic interaction with harmonic time dependence we show that
differences in the corresponding effective S--matrices do arise beyond
first--order perturbation theory.
We attribute them to the fact that the use of the direct reduction effective
Hamiltonian involves the additional approximation of neglecting contributions
from the negative--energy intermediate states, an approximation which is
unnecessary in the Foldy--Wouthuysen case as there the $4\times 4$
Hamiltonian does not connect positive-- and negative--energy states.
We conclude that at least in the cases where the relativistic Hamiltonian
is known, using the direct Pauli reduction effective Hamiltonian
introduces spurious relativistic effects and therefore the Foldy--Wouthuysen
reduction should be preferred.
\end{abstract}

\section{Introduction}
\label{1}

There are many cases in low-- and intermediate--energy nuclear physics
where one is confronted with the problem of incorporating relativistic
effects in a nonrelativistic calculation.
Such a situation arises, for example, when a relativistic Hamiltonian
 description exists for one part of the interaction but another part
can be realistically described  only by a nonrelativistic Hamiltonian.
Naturally, as the energies involved in these processes increase, inclusion
of relativistic corrections is expected to become more and more important.

There are at least two conceivable ways to construct ``effective $2\times 2$
Hamiltonians'' in order to identify relativistic corrections to nonrelativistic
Hamiltonians.

If the full relativistic Hamiltonian, $H$, is known, the Foldy--Wouthuysen (FW)
method \cite{Foldy,Bjorken,Itzykson,Barnhill,Friar,Hyuga,Ohta1,Goeller}
provides a systematic procedure to block--diagonalize the Hamiltonian order by
order in $1/M$ and hence to decouple
positive-- and negative--energy states to any desired order in $1/M$.
This is achieved by an application of successive unitary (in general
time--dependent) transformations on the operator $H-i\partial /\partial t$.
This method has frequently been used in the context of electromagnetic
processes in hadronic systems.

Alternatively, one can think of evaluating the matrix element of the
relativistic interaction Hamiltonian between free positive--energy solutions.
After reducing this matrix element to two--component form one may expand
the resulting expression to any order in $1/M$ and use it as an
effective Hamiltonian $H^{eff-P}_I$ to be evaluated between nonrelativistic
Pauli wave functions.
We shall refer to that procedure as the ``direct Pauli reduction''.

The purpose of this paper is to address the question
whether these two procedures yield identical results, and if not,
which of the two should be preferred, as they both are
used in the literature.
For example, in electron scattering from nuclei,
one evaluates matrix elements of
electromagnetic operators (``semi-de-relativized'' according to
the direct Pauli reduction procedure) between nonrelativistic
solutions of the Schr\"odinger equation with a
given $NN$ potential~\cite{Dubach,Fabian,Mathiot,Alberico,Hadjimichael},
but there also exist treatments that make use of the Foldy--Wouthuysen
procedure to obtain semi--relativistic operators~\cite{Goeller,Friar77,Ohta2}.
Similarly, in proton--proton bremsstrahlung the strong interaction part is
conventionally treated using a nonrelativistic Lippmann--Schwinger approach,
whereas for the electromagnetic interaction a semi--relativistic ansatz,
involving either the Foldy--Wouthuysen method  or the direct
Pauli reduction  method is
made \cite{Workman,Nakayama,Jetter,Brown,Katsog}.

There are basically two observations that provide the key to our investigation.
The first observation is that  there should be a clear distinction
between Hamiltonians that are obtained starting from a relativistic one
by means of unitary transformations (and which are still $4\times 4$ operators)
and what we shall refer to as effective $2\times2$ Hamiltonians (which are
$2\times 2$ operators, constructed as the upper left--hand parts of the
corresponding $4\times 4$ Hamiltonians).
As we shall show, the direct Pauli reduction can be also viewed as an effective
$2\times 2$ Hamiltonian obtained from a $4\times 4$ unitary time--independent
transformation of the relativistic Hamiltonian, by eliminating  the
contribution of the negative--energy sector of the relativistic Hamiltonian.
This distinction extends to the S--matrices also, where the $4\times 4$
Hamiltonians lead to the full relativistic S--matrix, whereas the
effective $2\times 2$ Hamiltonians lead to effective S--matrices.
Such effective S--matrices, which are calculated in nonrelativistic
scattering theory, may, depending on the interaction,
show effects of the neglect of the negative--energy sector.

The second observation is that in most practical applications
both these methods for obtaining effective Hamiltonians
are not applied to the {\it total} (that is, electroweak--plus--strong)
Hamiltonian but only to the electroweak part,
in combination with a nonrelativistic approach to the strong interaction.
Physical observables, such as S--matrix, are then calculated in the framework
of old--fashioned time--ordered perturbation theory.
The examples presented above, namely, electron scattering from nuclei and
proton--proton bremsstrahlung are clearly of that type.

It is commonly believed that both types of effective Hamiltonians lead to
identical effective S--matrices (and hence to identical observables)
and thus that it makes no difference which is used in a nonrelativistic
calculation.
A justification for this assumption has been given in
refs.~\cite{Nieto,Goldman}.
We discuss this assumption in quite some detail and find that {\em in general}
it is not correct. In particular, in the cases where the full relativistic
Hamiltonian is known, we show that the transformed
 $4\times 4$ Hamiltonians all lead to the same S--matrix, which is
(to the given order in $1/M$) identical to the relativistic S--matrix.
However, the {\it effective}  S--matrices corresponding to the
Foldy--Wouthuysen and the  direct Pauli reduction,
which are those appropriate for a nonrelativistic calculation,
differ from
each other beyond first order in perturbation theory and only the
Foldy--Wouthuysen effective S--matrix reproduces the full relativistic
S--matrix to that particular order in $1/M$. We discuss these issues
in detail for the particular case of Compton scattering by a proton
in ref.~\cite{paper2}.
In the cases where the full relativistic Hamiltonian is not known
we cannot argue in favour of any of the different effective S--matrices
in particular. In order to get a feeling about the importance
of these differences, we  compare the results of a
proton--proton bremsstrahlung calculation using the two different methods.

Our paper is organized as follows.
In the next section we  discuss general time--dependent unitary
transformations.
In this context we are mainly concerned with the Foldy--Wouthuysen
transformation.
In the third section we  compare
the first--order matrix elements of the effective Hamiltonians
obtained with the Foldy--Wouthuysen and direct Pauli reduction procedures
and investigate under which circumstances they lead to the same result.
In the fourth section we argue about the need to
go beyond first--order perturbation theory. Using a generic Hamiltonian with
the sole assumption of harmonic time dependence
we explain how differences arise in the different effective S--matrices.
In  section 5 we give a general nonperturbative  proof of why the S--matrix
elements corresponding to the $4\times 4$ transformed Hamiltonians are
identical with the full relativistic S--matrix.
We conclude and summarize our main results in section 6.

\section{The Foldy--Wouthuysen Transformation}

In this section we discuss time--dependent unitary transformations
with special emphasis on the Foldy--Wouthuysen transformation.

\subsection{Unitary Transformations of the Schr\"odinger Equation}
\label{2.1}
If we start with an equation of motion of the Schr\"odinger
type\footnote{We regard the Dirac equation as a specific example of a
Schr\"odinger type equation.},
\begin{equation}
\label{schroedinger}
i \frac{\partial |\Psi(t)>}{\partial t} = H(t) |\Psi(t)>,
\end{equation}
where we allow for an explicit time dependence of the Hamiltonian operator
$H(t)$, a unitary transformation $T(t)$
\begin{eqnarray}
\label{ut}
|\Psi'(t)> & = & T(t) |\Psi(t)>, \nonumber \\
T(t) T^{\dagger}(t) & = & T^{\dagger}(t) T(t) =1,
\end{eqnarray}
will result in the new Schr\"odinger equation
\begin{equation}
\label{newschroedinger}
i \frac{\partial |\Psi'(t)>}{\partial t} = \left( T(t) H(t) T^{\dagger}(t)
-i T(t) \frac{\partial T^{\dagger}(t)}{\partial t} \right) |\Psi'(t)>
\equiv H'(t) |\Psi'(t)>.
\end{equation}
In many applications, such as e.~g.~the Foldy--Wouthuysen transformation
\cite{Foldy,Bjorken,Itzykson}, it is useful to parameterize $T(t)$ as
\begin{equation}
\label{part}
T(t)=e^{i S(t)}, \quad S(t)=S^{\dagger}(t),
\end{equation}
and then to expand the new Hamiltonian $H'(t)$ in terms of $S(t)$,
\begin{eqnarray}
\label{newschroedingerexp}
H'(t) & = & H + i [S,H] - \frac{1}{2}[S,[S,H]] - \frac{i}{6}[S,[S,[S,H]]]
+ \dots \nonumber \\
& & -\dot{S} - \frac{i}{2}[S,\dot{S}] + \frac{1}{6}[S,[S,\dot{S}]]
+ \dots,
\end{eqnarray}
with $\dot{S}=\partial S/ \partial t$.

As the unitary transformation $T(t)$ may in general be time--dependent,
one clearly finds \cite{Itzykson,Nieto,Goldman}
\begin{equation}
\label{ineq}
<\Psi(t)|H(t)|\Psi(t)> \not= <\Psi'(t)|H'(t)|\Psi'(t)>,
\end{equation}
which simply expresses the fact that if $H(t)$ is the operator corresponding
to a physical observable in the first representation, $H'(t)$ is {\em not}
the corresponding operator describing the {\em same} observable in the second
representation \cite{Nieto,Goldman}. In other words, as was pointed out by
Nieto \cite{Nieto}, it is the unitary transform of $H(t)$,
i.~e.~$T(t) H(t) T^{\dagger}(t)$, which is physically
equivalent to $H(t)$, as it is this operator which
yields the same matrix element between transformed states as $H(t)$ between
the original states.

\subsection{The Foldy--Wouthuysen Method}
\label{2.2}
The Foldy--Wouthuysen transformation \cite{Foldy,Bjorken,Itzykson} provides a
systematic method of finding a representation of the Dirac Hamiltonian in which
positive-- and negative--energy states are separately represented by
two--component wave functions instead of the four--component wave functions
in the ordinary Dirac representation. For the free Dirac
equation\footnote{States and Hamiltonians without specific superscript labels
will always be assumed to be in the Dirac representation.}
\begin{equation}
\label{freediraceq}
i \frac{\partial \Psi_0 (x)}{\partial t}
= (\vec{\alpha} \cdot \vec{p} + \beta M) \Psi_0 (x) = H_0 \Psi_0 (x),
\end{equation}
where $\alpha_i$ and $\beta$ are the usual Dirac matrices, $\vec{p}$ is
the momentum operator, and $x$ is a shorthand notation for $(\vec{x},t)$,
the transformation is exactly known and given by
\begin{eqnarray}
\label{freefwtrafo}
\Psi^{FW}_0(x) & = & T_0 \Psi_0(x), \nonumber \\
T_0 & = & \sqrt{\frac{E+M}{2E}} \left( \begin{array}{cc}
1 & \frac{\vec{\sigma} \cdot \vec{p}}{E+M}\\
- \frac{\vec{\sigma} \cdot \vec{p}}{E+M} & 1
\end{array} \right),
\end{eqnarray}
where we have defined the {\em operator} $E=\sqrt{\vec{p}\,^2+M^2}$.
In the new Foldy--Wouthuysen representation the free Dirac equation has
the simple form
\begin{equation}
\label{fdefw}
i \frac{\partial \Psi^{FW}_0(x)}{\partial t}
= \beta \sqrt{\vec{p}\,^2+M^2}\, \Psi^{FW}_0(x) = H_0^{FW} \Psi^{FW}_0(x).
\end{equation}
As $\beta$ is of block--diagonal form, eq.~(\ref{fdefw}) is just the direct
sum of two Hamiltonians $\pm \sqrt{\vec{p}\,^2+M^2}$. The
positive/negative--energy solutions of eq.~(\ref{fdefw}) are of the form
\begin{eqnarray}
\label{fwsol}
\Psi^{FW(+)}_0(x) & = & \left ( \begin{array}{c} \chi^{(+)}(x) \\0
\end{array} \right ), \nonumber \\
\Psi^{FW(-)}_0(x) & = & \left( \begin{array}{c} 0 \\ \chi^{(-)}(x)
\end{array} \right),
\end{eqnarray}
where $\chi^{\pm}(x)$ and $0$ are two--component spinors.
Note that $H^{FW}_0$ is simply given by $T_0 H_0 T_0^{\dagger}$
because $T_0$ is time--independent.

If the Dirac Hamiltonian contains an explicitly time--dependent interaction
$H_I(t)$, the transformation $T$ will depend on time, $T=T(t)$, and,
in general, a closed form for the transformation $T(t)$ leading to a
block--diagonal form of the Hamiltonian is not known.
However, Foldy and Wouthuysen \cite{Foldy} developed a systematic procedure
to construct a new Hamiltonian which is block--diagonal to any desired order in
$1/M$.
The idea is to split the Dirac Hamiltonian into its odd and even
components\footnote{Odd operators ${\cal O}$, such as e.~g.~$\alpha_i$,
couple large and small components whereas even operators ${\cal E}$
(e.~g.~$\beta$) do not.
The following identities are useful in the derivation of the Foldy--Wouthuysen
transformation: $[{\cal O},\beta]=2{\cal O} \beta, [{\cal E},\beta]=0 $.},
\begin{equation}
\label{hoe}
H(t) = H_0 + H_I(t) = \beta M + {\cal O}(t) + {\cal E}_I(t),
\end{equation}
where ${\cal O}(t)={\cal O}_0 + {\cal O}_I(t)
= \vec{\alpha} \cdot \vec{p} + {\cal O}_I(t)$.  It is then assumed, that the
interaction potentials ${\cal O}_I$ and ${\cal E}_I$ do not contain
powers of $1/M$ that are smaller than $(1/M)^0$.
Furthermore, the interaction has to be {\em weak} enough in the sense
that the magnitude of each of the time and space Fourier components of the
interaction potential is considerably smaller than the mass of the
nucleon \cite{Foldy}.
It is then understood that we mean by $[1/M]$ ``terms of order
$E_{ref}/M$'', where $E_{ref}$ is some reference energy smaller than $M$.
The procedure consists
of first applying the transformation
\begin{eqnarray}
\label{trafo1}
T^{(1)}(t) & = & e^{i S^{(1)}(t)}, \nonumber\\
S^{(1)}(t) & = & - i \beta \frac{{\cal O}(t)}{2 M},
\end{eqnarray}
to eq.~(\ref{hoe}), the result of which is then written as
\begin{equation}
\label{hoe1}
H^{FW(1)}(t) = \beta M + {\cal O}^{(1)}(t) + {\cal E}^{(1)}(t).
\end{equation}
Using eq.~(\ref{newschroedingerexp}) it can be easily shown
\cite{Foldy,Bjorken,Itzykson} that the transformation of eq.~(\ref{trafo1})
is constructed such that the odd operator of eq.~(\ref{hoe1}) is of order
$1/M$. The procedure is then repeated with a new transformation
$T^{(2)}(t)$ which is exactly of the same form as
eq.~(\ref{trafo1}) except that in $S^{(2)}(t)$ the new odd operator
${\cal O}^{(1)}(t)$ of eq.~(\ref{hoe1}) appears.
As each successive application reduces the leading power of the
odd operator of the resulting Hamiltonian
by one unit, after applying this method n times one obtains
a Hamiltonian which is block--diagonal to order $1/M^{n-1}$.
For example, after four transformations the
Hamiltonian which does not contain any odd operators up to and including
order $1/M^3$ reads
\begin{eqnarray}
\label{hfw4}
H^{FW(4)}(t) &= & \beta \left( M + \frac{{\cal O}^2}{2 M} - \frac{{\cal O}^4}{8
M^3}
              \right )  + {\cal E}_I
- \frac{1}{8 M^2} [{\cal O},[{\cal O},{\cal E}_I] \nonumber \\ &&
+i \dot{{\cal O}_I}]
-\frac{\beta}{8 M^3} \left ( [{\cal O},{\cal E}_I] + i \dot{{\cal O}_I}
\right)^2 + [\frac{1}{M^4}],
\end{eqnarray}
where $\dot{{\cal O}_I}$ stands for $\partial {\cal O}_I/\partial t$.

\section{First--Order Matrix Elements}
It is in general not possible to explicitly solve the Dirac equation with an
arbitrary time--dependent potential. On the other hand in many cases
it is sufficient to construct a nonrelativistic approximation including
lowest--order relativistic corrections.
Such an approximation is often formulated in terms of a $2 \times 2$
nonrelativistic  effective Hamiltonian $H^{eff}$ which is to be used
in nonrelativistic perturbation theory.
One way of constructing such an effective Hamiltonian to be used in
low-- and intermediate--energy applications is to evaluate the
matrix element of the potential operator between positive--energy solutions
of the {\em free} Dirac equation, perform a two--component reduction  and then
interpret the
result as the matrix element of an effective $2\times2$ Hamiltonian between
nonrelativistic (positive--energy) states (direct Pauli reduction).
Commonly the so constructed
effective Hamiltonian is expanded in a series in $1/M$ and only the
terms to a particular order are kept.

\subsection{Reduction of the First--Order Matrix Element}
\label{3.1}
To be specific we consider the expression
\begin{equation}
\label{me}
\int d^3 x \, \Psi^{ \dagger}_{0f}(\vec{x},t) H_I(t)
\Psi_{0i}(\vec{x},t)
= \int d^3 x \, \Psi^{FW \dagger}_{0f} (\vec{x},t) T_0 H_I(t) T_0^{\dagger}
\Psi^{FW}_{0i} (\vec{x},t),
\end{equation}
where $\Psi_{0i}$ and $\Psi_{0f}$ are {\em positive}--energy solutions of
the free Dirac equation and $\Psi^{FW}_{0i}$ and $\Psi^{FW}_{0f}$ are the
corresponding solutions of the Foldy--Wouthuysen Hamiltonian
(see eq.~(\ref{freefwtrafo}) and
(\ref{fdefw})). Expressions of the type of eq.~(\ref{me}) typically appear
as building blocks of a perturbative treatment of the S--matrix.
{}From eq.~(\ref{me}) we find that instead of explicitly reducing the matrix
element, we may as well look at the operator $H^P_I = T_0 H_I(t)
T_0^{\dagger}$.
This operator, when taken between the states $\Psi^{FW}_{0i}$ and
$\Psi^{FW}_{0f}$,
leads to an effective $2 \times 2$ Hamiltonian $H^{eff-P}_I$ which is just what
one would get by performing a two--component Pauli reduction on the left--hand
side of eq.~(\ref{me}).

For a general interaction, $H_I(t)=\OI (t) + \EI (t)$, a reduction through
order $1/M^3$ yields
\begin{eqnarray}
\label{redtht}
H^P_I(t) & = & \EI + \frac{\beta}{2 M} \{ \OZ, \OI \}
-\frac{1}{8 M^2} [\OZ,[\OZ, \EI] ]
-\frac{\beta}{8 M^3} \{ \OZ^2, \{ \OZ, \OI \} \}
\nonumber \\ & &
-\frac{\beta}{16 M^3} [ \OZ^2, [\OZ, \OI]]
+ [1/M^4]+ \: \mbox{odd terms}.
\end{eqnarray}
Here the terms written explicitly are all even operators.
The odd terms do not contribute when eq.~(\ref{redtht})
is evaluated between free positive--energy states of the Foldy--Wouthuysen
representation.
However, as it is only these states which are considered in a nonrelativistic
calculation, one can regard the upper left--hand block of
$H^P_I(t)$ as an effective Hamiltonian $H^{eff-P}_I(t)$ to be used with
two--component Pauli spinors.

\subsection{Comparison with the Foldy--Wouthuysen Hamiltonian}
\label{3.2}
The Foldy--Wouthuysen Hamiltonian of eq.~(\ref{hfw4}) may be written as
\begin{equation}
\label{hfwsum}
H^{FW(4)}=H^{FW}_0 + H^{FW}_1(t) + H^{FW}_2(t) + H^{FW}_3(t) + H^{FW}_4(t)
+[1/M^4],
\end{equation}
where the subscripts $i$ indicate the order in which the interaction
potentials $\OI$ and $\EI$ appear in the corresponding Hamiltonian.
The {\em linear} interaction Hamiltonian is given by
\begin{eqnarray}
\label{red}
H^{FW}_1(t) & = & \EI + \frac{\beta}{2 M} \{ \OZ, \OI \}
-\frac{1}{8 M^2} [\OZ,[\OZ,\EI]+i\dot{\OI}]
\nonumber \\ & &
-\frac{\beta}{8 M^3} \{\OZ^2,\{\OZ,\OI\}\}+[1/M^4].
\end{eqnarray}
Comparing eq.~(\ref{redtht}) and eq.~(\ref{red}) we see that the two operators
$H^{FW}_1(t)$ and $H^P_I(t)$ differ by
\begin{eqnarray}
\label{diff}
\Delta H_1(t) & = & H^{FW}_1(t) - H^P_I(t) \nonumber \\
         & = & -\frac{i}{8 M^2} [\OZ,\dot{\OI}]
               + \frac{\beta}{16 M^3} [\OZ^2,[\OZ,\OI]]
               + \mbox{odd terms} + [1/M^4] \nonumber \\
         & = & \frac{1}{8 M^2} \left([H_0^{FW}, [\OZ,\OI]]-i[\OZ,\dot{\OI}]
               \right) + \mbox{odd terms} + [1/M^4]. \nonumber \\ & &
\end{eqnarray}
In obtaining the last line of eq.~(\ref{diff}) we expanded eq.~(\ref{fdefw})
and made use of $\OZ^2=\vec{p}\,^2$.

It has often (tacitly) been assumed that $\Delta H_1(t)$ of eq.~(\ref{diff})
vanishes when evaluated between free positive--energy Foldy--Wouthuysen
states. In which case, in a nonrelativistic calculation to first order in the
interaction, it should make
no difference which of the two Hamiltonians $H^P_I(t)$ and
$H^{FW}_1(t)$ is used for the interaction. We will now investigate
under which circumstances this assumption is correct.

Most applications of the Foldy--Wouthuysen transformation
are concerned with either static (atomic physics, hydrogen atom)
or harmonically time--dependent external potentials.
If we assume the potential $\OI (t)$ to have a harmonic time dependence,
we find for the time derivative $\dot{\OI}=\mp i \omega \OI$, where the
upper/lower sign refers to the absorption/emission of a quantum of energy
$\omega$. In this case the matrix element of $\Delta H_1(t)$ reads
\begin{eqnarray}
\label{diffhtd}
\lefteqn{\int d^3 x \, \Psi^{FW \dagger}_{0f}(\vec{x},t) \Delta H_1(t)
\Psi^{FW}_{0i}(\vec{x},t)  = } \nonumber \\
& & \frac{1}{8 M^2} (E_f-E_i \mp \omega)
\int d^3 x \, \Psi^{FW \dagger}_{0f}(\vec{x},t) [\OZ,\OI]
\Psi^{FW}_{0i}(\vec{x},t) + [1/M^4]. \nonumber \\ & &
\end{eqnarray}
{}From eq.~(\ref{diffhtd}) it is seen, that for a potential with harmonic
time dependence, both Hamiltonians will yield the same first--order
matrix elements if the energies of the initial
and final state match with the absorption/emission
of a quantum of energy $\omega$, or to put it in a somewhat
sloppy way, if energy is ``conserved'' in the transition, or alternatively
if the
commutator $[\OZ,\OI]$ vanishes, which in general we do not expect to
happen. This observation has also been made in ref.~\cite{Ohta1}.

In second--order (old--fashioned) perturbation theory,
energy is not conserved at a vertex \cite{Weinberg,Halzen}.
This clearly suggests,
that in practical second--order calculations,
such as e.~g.~proton--proton bremsstrahlung, it may make a
difference whether one uses the linear effective Foldy--Wouthuysen Hamiltonian
$H_1^{eff-FW}(t)$ deduced from
eq.~(\ref{red}) or the Hamiltonian $H^{eff-P}_I(t)$ obtained from
eq.~(\ref{redtht}).

Our result seems to be somewhat different from the earlier work of Nieto
\cite{Nieto} and of Goldman \cite{Goldman}, who appear to have shown that the
results from the two first--order Hamiltonians are the same, i. e. that the
matrix element of $\Delta H_1(t) = 0$ in general. To understand this
difference,
consider again this matrix element. Using the last part of eq.~(\ref{diff}) one
can write
\begin{eqnarray}
\label{diffhtd2}
\lefteqn{\int d^3 x \, \Psi^{FW \dagger}_{0f}(\vec{x},t) \Delta H_1(t)
\Psi^{FW}_{0i}(\vec{x},t)  = } \nonumber \\
& & -
\int d^3 x \, \Psi^{FW \dagger}_{0f}(\vec{x},t)
\left([\Delta,H_0^{FW}]+i\dot{\Delta} \right)
\Psi^{FW}_{0i}(\vec{x},t) + [1/M^4],
\end{eqnarray}
where to the order being considered $\Delta = [\OZ,\OI]/8M^2$. A more general
form
for $\Delta$ is given in ref.~\cite{Goldman}, but is unnecessary for the
purposes of this discussion.

The authors of both ref.~\cite{Nieto} and ref.~\cite{Goldman} make the {\em
assumption}
that
\begin{eqnarray}
\label{schrst}
[\Delta,H_0^{FW}]+i\dot{\Delta} = 0,
\end{eqnarray}
which they refer to as the ``Schr\"odinger statement'', and thus conclude that
\mbox{$< \Psi^{FW \dagger}_{0f}(t) | \Delta H_1(t) |
\Psi^{FW}_{0i}(t) > \equiv 0$} and consequently that it makes no difference
which of
the Hamiltonians is used.

It is clear however from eq.~(\ref{diff}) that eq.~(\ref{schrst}) cannot be
correct in general as an operator equation. If one considers it instead as a
matrix
equation, then the Schr\"odinger equation can be used to replace the
Hamiltonian
by a time derivative and one obtains
\begin{eqnarray}
\label{schrst2}
\lefteqn{ <\Psi^{FW}_{0f}(t)| \left([\Delta,H_0^{FW}]+i\dot{\Delta} \right)|
\Psi^{FW}_{0i}(t)>  = } \nonumber \\ & &  i \frac{d}{dt} <\Psi^{FW}_{0f}(t)|
\Delta | \Psi^{FW}_{0i}(t)> + [1/M^4].
\end{eqnarray}
Again, however, the right--hand side of this equation is not identically zero
in
general. If one assumes a harmonic time dependence of the interaction, as we
have
done above, then the time dependence of the matrix element is just
$e^{i(E_f-E_i \mp \omega)t}$ and the time derivative simply brings down the
energy factor $i(E_f-E_i \mp \omega)$. This leads exactly to our previous
eq.~(\ref{diffhtd}) and to our previous conclusion that the matrix element of
$\Delta H_1$ vanishes, and the results using different Hamiltonians are
equivalent, only when the transition conserves energy. In the more general case
when the states are off--energy--shell, as will be the case when such matrix
elements are used in second--order perturbation theory, different Hamiltonians
may be expected to give different results.

As a specific example of a case where the choice of Hamiltonian does make a
difference, let us consider the interaction of a proton with an
external electromagnetic field,
\begin{eqnarray}
\label{emint}
\EI(t) & = & e \Phi(\vec{x},t)
             -\frac{e \kappa}{2 M} \beta \vec{\sigma} \cdot \vec{B}(\vec{x},t),
\nonumber \\
\OI(t) & = &  -e \vec{\alpha} \cdot \vec{A}(\vec{x},t)
              + i \frac{e \kappa}{2 M} \beta \vec{\alpha} \cdot
\vec{E}(\vec{x},t),
\end{eqnarray}
where $e>0$ is the elementary charge, $\kappa=1.79$ the anomalous magnetic
moment in units of a nuclear magneton, and $\vec{E}=-\vec{\nabla} \Phi
- \dot{\vec{A}}$ and $\vec{B} = \vec{\nabla} \times \vec{A}$.
Using this specific example the {\em even} part of
$\Delta H_1(t)$ is given to order $1/M^3$ by
\begin{eqnarray}
\label{diffem}
\Delta H_1(t) & = & - \frac{e \beta}{16 M^3} [\vec{p}\,^2,-i \vec{\nabla}
\cdot \vec{A} + i \vec{\sigma} \cdot \vec{p} \times \vec{A}
-i \vec{\sigma} \cdot \vec{A} \times \vec{p}] \nonumber \\
& & -\frac{e \kappa}{16 M^3} \beta ( \vec{p} \cdot \dot{\vec{E}}
+ \dot{\vec{E}} \cdot \vec{p} + \vec{\sigma} \cdot \vec{\nabla}
\times \dot{\vec{E}} ) \nonumber \\
& & + \frac{ i e }{8 M^2} ( - i \vec{\nabla} \cdot \dot{\vec{A}}
+ i \vec{\sigma} \cdot \vec{p} \times \dot{\vec{A}}
- i \vec{\sigma} \cdot \dot{\vec{A}} \times \vec{p} ).
\end{eqnarray}
In all modern proton--proton bremsstrahlung calculations
\cite{Workman,Nakayama,Jetter,Brown,Katsog}
the electromagnetic part of the interaction is obtained by reducing the
relativistic nucleon--nucleon--gamma vertex to obtain an effective $2 \times 2$
Hamiltonian
which is used with two--component spinors in a nonrelativistic calculation. In
some cases, e.~g.~ref.~\cite{Workman}, the effective Hamiltonian
$H^{eff-FW}_1$ obtained via a
Foldy--Wouthuysen reduction as in eq.~(\ref{red})  was used and in others,
e.~g.~\cite{Nakayama},
the $H^{eff-P}_I$ obtained via a direct two--component reduction as in
eq.~(\ref{redtht}) was used. These two should differ by the $\Delta H_1$ of
eq.~(\ref{diff}) or eq.~(\ref{diffem}).

In figure 1 the results of such a bremsstrahlung calculation
are shown for a particular kinematic
situation, corresponding to a coplanar equal--angle geometry with outgoing
protons measured at $\Theta_3 = \Theta_4 = 6^{\circ}$ and with an
incident laboratory energy $T^{LAB} = 280$ MeV. The calculations of
ref.~\cite{Workman} have been used taking a) the
Foldy--Wouthuysen Hamiltonian for the electromagnetic interaction
of the proton or b) the effective Hamiltonian
of eq.~(\ref{redtht}). As can be seen from eq.~(\ref{diffhtd})
the difference between the two calculations is of order
$1/M^2$ multiplied by the amount by which energy is not
conserved at the electromagnetic vertex. For the kinematics
chosen in figure 1
the difference is of order 7 \%.

We conclude that as long as one considers only first--order matrix elements,
such as those appearing in $\mu + p \rightarrow n + \nu$ or electron scattering
from an on--shell nucleon \cite{McVoy}
it does not make a difference which reduction scheme is used.
This is true because in a perturbative treatment of the S--matrix
the time integration enforces energy conservation at the vertex
of lowest--order perturbation theory.
However, as we have seen in figure 1 the situation is different for
higher--order processes such as e.~g.~proton--proton bremsstrahlung.

One might properly fault this particular numerical example because the strong
interaction is not treated consistently with the electromagnetic interaction,
i.~e.~not as a reduction of some relativistic interaction. Thus it is important
to look at more detail at second--order processes in which both interactions
can
be treated in the same way. That we do in general in the
following section and for the specific case of Compton scattering by a proton
in
ref.~\cite{paper2}.

\section{Second--Order Perturbation Theory }

In the previous sections we described how to obtain various effective
interaction Hamiltonians for use in nonrelativistic calculations and
showed that the first--order matrix elements of these Hamiltonians differed
only in situations when the states involved did not conserve energy.
Such a situation arises in second-- (or higher--) order old--fashioned
time--ordered perturbation theory where such first--order matrix
elements connect to an intermediate state.
Of course, the {\em total} energy is conserved for each (higher--order)
diagram,
but energy is not conserved at the individual vertices of such a diagram
\cite{Weinberg,Halzen}.
It is this fact which {\em may} lead to different predictions for the
effective S--matrix
and thus it is important to see what happens in such second--order processes.

There are two questions which have to be addressed.
Firstly, which kind of transformation of the $4\times4$ Hamiltonian
leads to the same relativistic S--matrix?
As long as we are dealing with Hamiltonians originating from a unitary
transformation of the states (see section 2.1) one would expect that the full
relativistic S--matrix would be invariant under such transformations.
That is indeed the case as we will show in second--order perturbation theory in
the first part of this section and by more formal manipulations valid to all
orders in sec.~\ref{5}.

This invariance of the full relativistic S--matrix has been observed by others,
and is consistent with our usual understanding of quantum mechanics.
However it has led to the extremely misleading, and incorrect, assumption that
the physical observables calculated with various {\em effective} Hamiltonians
will be the same and thus that it makes no difference which effective
Hamiltonian is used.

Hence we are led to the second question which is of great practical importance.
Do different effective $2\times2$ Hamiltonians lead to the same nonrelativistic
effective S--matrix?
We show in the last part of this section that in fact this is not the case.
The reason is that the construction of effective Hamiltonians is not always
done solely via a unitary transformation of the states.
Thus it {\em does} make a difference which effective Hamiltonian
is used in nonrelativistic calculations.
Only the Foldy--Wouthuysen Hamiltonian (and other block--diagonal Hamiltonians
obtained from it) reproduces the full relativistic S--matrix to a given order
in $1/M$.
Thus it is only such Hamiltonians which can be presumed to give correct
physical results in nonrelativistic calculations.

\subsection{Full Relativistic S--Matrix in Second--Order Perturbation Theory}
\label{4.1}

We now restrict our consideration to second--order perturbation theory. Thus we
assume that the relativistic $4 \times 4$ Dirac Hamiltonian is given as before
by $H
= H_0 + H_I(t)$, where $H_0$ is the free Dirac Hamiltonian. To be definite we
will
focus on processes involving an incoming quantum of energy $\omega_a$
and an outgoing one of energy $\omega_b$. Thus $H_I(t)$ will have a term $ \sim
H_a
e^{-i \omega_a t}$ and one $ \sim H_b e^{i \omega_b t}$as well as others as
required for Hermiticity. Formally we could construct $H_I(t)$ in second
quantized form and use appropriate creation and destruction operators to pick
out the pieces needed. Instead we will treat $H_I(t)$ somewhat loosely with the
understanding that, at the end, by second--order we will mean that we will
discard
all terms except those containing one power of the coupling associated with
$H_a$ and one power of that associated with $H_b$.
It may be that $H_a = H_b$, as for example in Compton scattering where $ H_a =
H_b \sim H_{\gamma N N}$. Alternatively $H_a$ may be different from $H_b$ as in
pion photoproduction where we could take $H_a \sim H_{\gamma N N}$ and $H_b
\sim H_{\pi N N}$.

With these assumptions the full relativistic S--matrix is given through second
order in the interaction in the standard
fashion, e.~g.~ref.~\cite{Bjorken}, eq.~(6.57), as
\begin{eqnarray}
\label{relsmatrix}
S_{fi} & = & \delta_{fi} -i \int d^4y \Psi^\dagger_{0f}(y) H_I(y) \Psi_{0i}(y)
\nonumber \\
& & -i \int d^4y d^4z \Psi^\dagger_{0f}(y) H_I(y) S_F(y-z) \beta H_I(z)
\Psi_{0i}(z),
\end{eqnarray}
where $\Psi_{0f}$ and $\Psi_{0i}$ are {\em positive}--energy eigenstates of the
free Dirac equation and where $S_F$ is
the free Feynman propagator. To proceed further we need to examine how the
states, Hamiltonian, and propagator transform under the unitary transformation
$T(t)$.

We define the states $\Psi_{0p}(x)$ to be eigenstates of the free Dirac
Hamiltonian $H_0$.
They may have positive or negative  energy and can be written as
\begin{eqnarray}
\label{states1}
\Psi_{0p}^{(\pm)}(\vec{x},t) & = & e^{\mp i E_p t} \Phi_{0p}^{(\pm)}(\vec{x}),
\end{eqnarray}
where $E_p = +\sqrt{\vec{p}\,^2 + m^2}$ and
\begin{eqnarray}
\label{states2}
\Phi_{0p}^{(\pm)}(\vec{x}) &=& w^{(\pm)}(p) \frac{e^{\pm i \vec{p} \cdot
\vec{x}}}{(2\pi)^{\frac{3}{2}}},
\end{eqnarray}
where $w^{(+)} = u $  and $w^{(-)} = v$ with $u^\dagger u = v^\dagger v = 1$
are the usual Dirac spinors\footnote{Note that we use a different
normalization convention in comparison with ref.~\cite{Bjorken}.}.
Spins will always be summed, so all explicit spin
dependence will be suppressed.
These states satisfy a completeness relation
\begin{eqnarray}
\label{completeness}
\sum_{\rm spins} \int d^3p \left \{
\Phi_{0p}^{(+)}(\vec{x})\Phi_{0p}^{(+)\dagger}(\vec{y}) +
 \Phi_{0p}^{(-)}(\vec{x})\Phi_{0p}^{(-)\dagger}(\vec{y}) \right \} & = &
\delta^3(\vec{x}-\vec{y}).
\end{eqnarray}

Under the unitary transformation  $T_0$, which is all that will be needed, we
have $\Psi_{0p}^{FW (\pm)} (x) = T_0 \Psi_{0p}^{(\pm)} (x) $. This changes only
the
spinors which become, using the specific $T_0$ of eq.~(\ref{freefwtrafo}),
\begin{equation}
\label{spinors}
u(p) \rightarrow \left ( \begin{array}{c} \chi \\0 \end{array} \right )
\; \; \; {\rm and} \; \; \;
v(p) \rightarrow \left ( \begin{array}{c} 0 \\ \chi \end{array} \right ),
\end{equation}
with $\chi$ a two--component spinor and $0$ a two--component null spinor. We
will
always assume that $T_0$ is obtained from $T(t)$ by turning off the
interaction,
which thus removes the time dependence. For most of this section however it is
not necessary to specify $T_0$ or $T(t)$ explicitly. Thus we will use
$\Psi^{\prime}$ and $H^{\prime}$ for example for states and Hamiltonians
obtained via general transformations as in sec.~\ref{2.1}. We will reserve
$\Psi^{FW}$ and $H^{FW}$ for situations where it is important that the states
have the form eq.~(\ref{spinors}) and that the Hamiltonian is block--diagonal
to some
order in $1/M$, and thus where we require the specific transformation $T_0$ of
eq.~(\ref{freefwtrafo}) and the specific $T(t)$ of sec.~\ref{2.2}.

To determine how the propagator $S_F$ transforms it is easiest to first expand
in the complete set of states just defined. Thus starting with,
e.~g.~eq.~(6.48)
of ref.~\cite{Bjorken}, and using a standard expansion for $\theta(t-t')$ we
obtain
\begin{eqnarray}
\label{SFexpan}
S_F(y-z)\beta & = &  \frac{1}{2 \pi} \int dp_0 d^3p  e^{-i
p_0 (t_y-t_z)}
\sum_{\rm spins} \left \{ e^{-iE_p (t_y-t_z)} \frac{  \Phi^{(+)}_{0p}(\vec{y})
\Phi^{(+)\dagger}_{0p}(\vec{z}) }{p_0+i\epsilon}  \right. \nonumber \\
& & \left. \mbox{} - e^{iE_p (t_y-t_z)} \frac{ \Phi^{(-)}_{0p}(\vec{y})
\Phi^{(-)\dagger}_{0p}(\vec{z})}{-p_0+i\epsilon}
 \right \}.
\end{eqnarray}
We now define
\begin{equation}
\label{SFprime}
S'_F(y-z)\beta = T_0 S_F(y-z)\beta T_0^\dagger.
\end{equation}
It is clear by applying $T_0 \ldots{} T_0^\dagger$ to the right--hand side of
eq.~(\ref{SFexpan}) that \mbox{$S'_F(x-y)\beta$} has exactly the same form as
$S_F(x-y)\beta$, only with the states $\Phi$ replaced by $\Phi'$ and so is
the correct free Feynman propagator in the $\Phi'$ representation.

Now consider the Hamiltonian. To allow an expansion to second order we write
the transformation variously as
\begin{equation}
\label{Texpan}
T(t) = e^{iS} = T_0 e^{i\tilde{S}} = T_0 e^{i(S_1 + S_2 + \ldots)},
\end{equation}
where $S_n$ is Hermitian and of order n in the interaction.
Starting from $H' = T(t) H(t) T^{\dagger}(t)
-i T(t) \dot{T}^{\dagger}(t)$ and inverting, we find for $H$
\begin{eqnarray}
\label{Heqn1}
H & = & e^{-i\tilde{S}} \left \{ T_0^\dagger H' T_0 +i e^{i\tilde{S}}
\frac{\partial
e^{-i\tilde{S}}}{\partial t} \right \}
e^{i\tilde{S}}.
\end{eqnarray}
We now write $H'$ as $H' = H'_0 + H'_1 + H'_2 +\ldots{}$, where $H'_n$ is of
order $n$ in the interaction and expand $H$ in a fashion analogous to
eq.~(\ref{newschroedingerexp}) to obtain through second order
\begin{eqnarray}
\label{Heqn2}
H_I & = & \left \{ T_0^\dagger H'_1 T_0 - \Delta H^{\prime}_1 \right \} +
\left \{ T_0^\dagger H'_2 T_0 - \Delta H^{\prime}_2 \right \},
\end{eqnarray}
\begin{equation}
\label{dh1}
\Delta H^{\prime}_1  =  T_0^{\dagger}\Delta H_1 T_0  =  i[S_1,H_0] - \dot{S_1},
\end{equation}
\begin{eqnarray}
\label{dh2}
\Delta H^{\prime}_2 & = & i[S_2,H_0] - \dot{S_2} - \frac{i}{2} [S_1,\Delta
H^{\prime}_1] +
i[S_1,T_0^\dagger H'_1 T_0].
\end{eqnarray}
$H'_2$ is second--order in the interaction and is a contact term which is
generated by the transformation. $H_I$ is however first--order in the
interaction, so the second bracketed expression in eq.~(\ref{Heqn2}) must be
zero. Hence $\Delta H^{\prime}_2 =  T_0^\dagger H'_2 T_0$.

The next step is to substitute this result for $H_I$ into
eq.~(\ref{relsmatrix})
and use eq.~(\ref{SFprime}) for $S'_F(x-y)\beta$ to obtain a result for that
part of the full
relativistic S--matrix  which is second--order in the interaction,
\begin{eqnarray}
\label{relsmatrix2}
S_{fi} & = & -i \int d^4y \Psi'^{\dagger}_{0f}(y) H'_2(y) \Psi'_{0i}(y)
\nonumber \\
& & -i \int d^4y d^4z \Psi'^{\dagger}_{0f}(y) H'_1(y) S'_F(y-z) \beta H'_1(z)
\Psi'_{0i}(z) \nonumber \\ & & + \Delta S_{fi},
\end{eqnarray}
where
\begin{eqnarray}
\label{deltasfi1}
\Delta S_{fi} & = &  i \int d^4y \Psi^{\dagger}_{0f}(y) \Delta H^{\prime}_2(y)
\Psi_{0i}(y) \nonumber \\
& & -i \int d^4y d^4z \Psi^{\dagger}_{0f}(y)
\Big \{ -T_0^\dagger H'_1(y) T_0 S_F(y-z) \beta
\Delta H^{\prime}_1(z) \nonumber \\
& & - \Delta H^{\prime}_1(y)  S_F(y-z) \beta T_0^\dagger H'_1(y) T_0
\nonumber \\& &
+ \Delta H^{\prime}_1(y) S_F(y-z) \Delta H^{\prime}_1(z)
\Big \} \Psi_{0i}(z),
\end{eqnarray}

Observe now that if $\Delta S_{fi} = 0$, as we will in fact show, then the
right--hand side of eq.~(\ref{relsmatrix2}) is exactly the
S--matrix that one would compute to second--order perturbation theory using
the transformed Hamiltonian $H'$ and the transformed states.
Comparison with eq.~(\ref{relsmatrix}) shows then that the S--matrix is
invariant under the unitary transformations of the type we have considered.

Thus consider $\Delta S_{fi}$. It is sufficient to look only at the $\Delta H
\times \Delta H$ term as the others are all similar. This term becomes after
using
the explicit expression for $S_F$, eq.~(\ref{SFexpan}), and extracting the time
dependence from the interaction and the states

\begin{eqnarray}
\label{deltas1}
\lefteqn{  - \frac{i}{2 \pi} \int dt_y dt_z dp_0 d^3p  \times } \nonumber \\
& & \left [ \sum_{\rm spins} \left \{ e^{-iA_+t_y+iB_+t_z}
\frac{<\Phi_{0f} | \Delta
H^{\prime}_{1b} | \Phi^{(+)}_{0p}>
<\Phi^{(+)}_{0p} | \Delta H^{\prime}_{1a} | \Phi_{0i}>}{p_0+i\epsilon}
\right. \right. \nonumber
\\ & &  \left. \left. \mbox{} -
 e^{-iA_-t_y+iB_-t_z} \frac{<\Phi_{0f} | \Delta H^{\prime}_{1b}
| \Phi^{(-)}_{0p}>
<\Phi^{(-)}_{0p} | \Delta H^{\prime}_{1a} | \Phi_{0i}>}{-p_0+i\epsilon}
\right \} + \quad \mbox{c.~t.}   \right ] \nonumber \\  & = &
-2 \pi i \delta (E_f + \omega_b - E_i - \omega_a) \int d^3p \times \nonumber
\\ & & \left [ \sum_{\rm spins} \left \{
\frac{<\Phi_{0f} | \Delta H^{\prime}_{1b} | \Phi^{(+)}_{0p}>
<\Phi^{(+)}_{0p} | \Delta H^{\prime}_{1a} | \Phi_{0i}>}{E_f + \omega_b
-E_p+i\epsilon}
\right. \right. \nonumber \\ & & \left. \left. \mbox{} +
\frac{<\Phi_{0f} | \Delta H^{\prime}_{1b} | \Phi^{(-)}_{0p}>
<\Phi^{(-)}_{0p} | \Delta H^{\prime}_{1a} | \Phi_{0i}>}{E_f + \omega_b
+E_p-i\epsilon}
\right \} + \quad \mbox{c.~t.}  \right ],
\end{eqnarray}
where $A_\pm = p_0-E_f-\omega_b\pm E_p$ and $B_\pm = p_0-E_i-\omega_a\pm E_p$.
Here c.~t.~indicates that there is a standard cross term
which has not been written explicitly.

Using the time dependence of $S_1$, which is linear in $H_1$, the matrix
elements
of $\Delta H^{\prime}_1$ can be evaluated as
\begin{eqnarray}
\label{deltah1}
<\Phi^{(\pm)}_{0p} | \Delta H^{\prime}_{1a} | \Phi_{0i}> & = &
<\Phi^{(\pm)}_{0p} | i[S_{1a}, H_0] + i \omega_a S_{1a} | \Phi_{0i}> \nonumber
\\
& = & i(E_i + \omega_a \mp E_p) <\Phi^{(\pm)}_{0p} | S_{1a} | \Phi_{0i}>,
\end{eqnarray}
\begin{eqnarray}
\label{deltah2}
<\Phi_{0f} | \Delta H^{\prime}_{1b} | \Phi^{(\pm)}_{0p}> & = &
<\Phi_{0f} | i[S_{1b}, H_0] - i \omega_b S_{1b} | \Phi^{(\pm)}_{0p}> \nonumber
\\
& = & -i(E_f + \omega_b \mp E_p) <\Phi_{0f} | S_{1b} | \Phi^{(\pm)}_{0p}>.
\end{eqnarray}
These matrix elements are analogous to that of eq.~(\ref{diffhtd}) but, since
they connect to intermediate states $\Phi_{0p}^{(\pm)}$, energy is not
conserved
and the energy factors do not vanish.

Finally, using these equations in a symmetric way (so as to end up with the
commutator) and using the completeness
relation eq.~(\ref{completeness}), one finds for eq.~(\ref{deltas1})
\begin{equation}
\label{deltas2}
-2 \pi i \delta(E_f + \omega_b -E_i - \omega_a) <\Phi_{0f} |-  \frac{i}{2}
[S_1,
\Delta H^{\prime}_1] | \Phi_{0i}>.
\end{equation}
To simplify the notation, the formula has been rewritten with the full
interaction, thus reintroducing the $H_a \times H_a$ and $H_b \times H_b$ terms
which should in the end be dropped.

The other terms are evaluated in exactly the same way so that one obtains
\begin{eqnarray}
\label{deltasfi2}
\Delta S_{fi} & = & 2 \pi i \delta(E_f + \omega_b -E_i - \omega_a) <\Phi_{0f} |
\Delta H^{\prime}_2 + \frac{i}{2} [S_1,\Delta H^{\prime}_1]   \nonumber \\
& &    \mbox{}  -i[S_1,T_0^\dagger H'_1 T_0]    |\Phi_{0i}> \nonumber \\
& = & 2 \pi i \delta (E_f + \omega_b -E_i - \omega_a) <\Phi_{0f} |
i[S_2,H_0]-i(\omega_b-\omega_a) S_2 |\Phi_{0i}> \nonumber \\
& = & 2 \pi  \delta (E_f + \omega_b -E_i - \omega_a)(E_f + \omega_b -E_i -
\omega_a)
<\Phi_{0f} | S_2 |\Phi_{0i}> \nonumber \\
&= & 0.
\end{eqnarray}
Here the first reduction uses the definition of $\Delta H^{\prime}_2$ of
eq.~(\ref{dh2})
and the second uses overall conservation of energy.

In summary, we have in this section considered the full relativistic S--matrix
in second--order perturbation theory (though to all orders in $1/M$). For
definiteness we focussed only on those processes involving an incoming quantum
of energy $\omega_a$ and outgoing quantum of energy $\omega_b$. In that context
we could show explicitly that the $\Delta S_{fi}$ of eq.~(\ref{relsmatrix2})
was zero.
The remaining terms on the right--hand side of that equation give just the full
relativistic S--matrix expressed in terms of the transformed Hamiltonian $H'_I$
and the transformed states $\Psi'_0$ to second order.
Thus we have shown explicitly that the full relativistic S--matrix is unchanged
under these transformations in second--order perturbation theory, which is a
special case of the more general, but more formal, proof to be given below.

Note that nowhere in this proof did we use any of the properties of $T(t)$,
except those of eq.~(\ref{Texpan}). Thus the result that the full relativistic
S--matrix is invariant holds for the transformation $T(t)$ of sec.~\ref{2.2}
which block--diagonalizes $H_I$ and
produces the Foldy--Wouthuysen Hamiltonian $H^{FW}_I$. It also holds for $T(t)
=T_0$, where the specific $T_0$ of eq.~(\ref{freefwtrafo}) is meant, which
produces $H_I^P$.

One other observation is worth making. The matrix element of $\Delta H$ is
essentially
the difference between the matrix elements of $H_I$ in the $\Phi_0$ basis and
$H'_I$ in the $\Phi'_0$ basis. From the details of the proof that $\Delta
S_{fi}
= 0$, as outlined in eqs.~(\ref{deltas1}) - (\ref{deltasfi2}),
one can see that these $\Delta H$ terms cancel the energy denominators
and thus generate contact terms. In general these contact terms
which  make up $\Delta H_2$ came {\em both} from positive-- and
negative--energy intermediate states.
This is somewhat in contrast to the generally held
belief that it is the negative--energy intermediate states alone which are
responsible for contact terms. It agrees however with the specific result we
have obtained in ref.~\cite{paper2}. There the specific example of Compton
scattering by a proton is described in detail and we can also see at which
order
in $1/M$ the various terms arise.

\subsection{Comparison of Effective Nonrelativistic S--matrices in
Second--Order
Perturbation Theory}
\label{4.2}

We now want to return to the main question, namely whether or not different
reductions of the relativistic interaction to effective nonrelativistic
interactions lead in practical calculations to different results. To do this we
need first to review how these effective Hamiltonians are obtained and used.

Usually one starts with a supposedly known relativistic interaction and makes a
transformation on it followed by a projection onto the upper left--hand block
which leads to a $2 \times 2$ effective interaction Hamiltonian,
$H^{eff}_I$, which is then used together with positive--energy two--component
states to construct an effective S--matrix, $S^{eff}_{fi}$, according to the
usual
rules of nonrelativistic scattering theory. Terms of higher than leading order
in $1/M$
included in the effective Hamiltonian provide relativistic corrections to the
lowest order nonrelativistic result and one hopes that if enough terms are
included $S^{eff}_{fi}$ will approach the full relativistic, and thus
presumedly
correct, result, $S_{fi}$.

Thus the question which is relevant for practical calculations is whether or
not these effective S--matrices are the same for different reductions of the
relativistic Hamiltonian, i.~e.~, for different $H^{eff}_I$, and whether they
equal the full relativistic $S_{fi}$.

In the Foldy--Wouthuysen case the new $H^{FW}_I$ is obtained via a
transformation
$T(t)$ of the type we have been considering, constrained by the condition that
to some order in $1/M$ the $4 \times 4$ Hamiltonian $H^{FW}_I$ is
block--diagonal.
The $2 \times 2$ effective Hamiltonian $H^{eff-FW}_I$ is then just the upper
left--hand
block, i.~e.~that part which has nonzero matrix element between free
positive--energy states $\Phi^{FW}_0$. Note that by virtue of the fact that
$T(t)$ depends on
the interaction, $H^{eff-FW}_I$ will have terms of higher order in the
interaction,
even though $H_I$ was first--order.

An alternative approach often used is to make a direct Pauli reduction of the
relativistic matrix element $<\Phi_{0f} | H_I | \Phi_{0i} > = <\Phi^{FW}_{0f} |
T_0
H_I T_0^\dagger | \Phi^{FW}_{0i} >$.  Thus in this case the $4 \times 4$
Hamiltonian $H^P_I = T_0 H_I T_0^\dagger $ and the $2 \times 2$ $H^{eff-P}_I$
is just the upper left--hand block of this. Note that in this case $H^P_I$ is
not
block--diagonal and that $H^{eff-P}_I$ may or may not be expanded (or
truncated) in
powers of $1/M$. Furthermore $H^{eff-P}_I$ is linear in the interaction, so
there are no contact terms generated naturally.

To see how the effective S--matrices calculated from these effective
Hamiltonians are related to the full relativistic S--matrix we start with
eq.~(\ref{relsmatrix2}) above with $\Delta S_{fi} = 0$ and use
eq.~(\ref{SFexpan}) and eq.~(\ref{SFprime}) to expand the
propagator into positive-- and negative--energy states. The time dependence can
then be extracted and the formula simplified in exactly the same way the
$\Delta H
\times \Delta H$ term was treated. The result for the second--order
contributions only is
\begin{eqnarray}
\label{effsfi1}
S_{fi} & = & \mbox{} -2 \pi i \delta (E_f + \omega_b - E_i - \omega_a)
\Bigg\{  <\Phi^{FW}_{0f} | H'_2 | \Phi^{FW}_{0i}>
\nonumber \\ &  & + \sum_{\rm spins} \int d^3p
\frac{<\Phi^{FW}_{0f} | H'_{1b} | \Phi^{FW(+)}_{0p}>
<\Phi^{FW(+)}_{0p} | H'_{1a} | \Phi^{FW}_{0i}>}{E_f + \omega_b -E_p+i\epsilon}
\nonumber \\ & & \mbox{} +
\frac{<\Phi^{FW}_{0f} | H'_{1b} | \Phi^{FW(-)}_{0p}>
<\Phi^{FW(-)}_{0p} | H'_{1a} | \Phi^{FW}_{0i}>}{E_f + \omega_b +E_p-i\epsilon}
+ \quad \mbox{c.~t.}   \Bigg \}.
\nonumber \\ &&
\end{eqnarray}
The term with $H'_2$ and the terms with $ \Phi^{FW(+)}_{0p}$ involve only
positive--energy states,
which have only upper components. Hence these terms pick out just the
upper left--hand block of $H'_1$ and $H'_2$ and consequently make up just the
effective S--matrix
one would calculate using the $2 \times 2$ $H^{eff}_I$. Thus we can write
\begin{eqnarray}
\label{effsfi2}
S_{fi} & = & S^{eff}_{fi} -2 \pi i \delta (E_f + \omega_b - E_i - \omega_a)
\times \nonumber \\ & & \left \{\sum_{\rm spins} \int d^3p
\frac{<\Phi^{FW}_{0f} | H'_{1b} | \Phi^{FW(-)}_{0p}>
<\Phi^{FW(-)}_{0p} | H'_{1a} | \Phi^{FW}_{0i}>}{E_f + \omega_b +E_p-i\epsilon}
+ \quad \mbox{c.~t.} \right \}.\nonumber \\ & &
\end{eqnarray}

Now it is obvious that for the Foldy--Wouthuysen reduction
the last term vanishes, since there $H^{FW}_I$
is block--diagonal and since a block--diagonal $H^{FW}_I$ does
not connect positive-- and negative--energy states. Thus in this case $S_{fi} =
S_{fi}^{eff-FW}$ to some order in $1/M$. Hence the Foldy--Wouthuysen procedure
gives the correct effective S--matrix when $H^{eff-FW}_I$ is used as a $2
\times
2$ nonrelativistic Hamiltonian in standard nonrelativistic scattering theory.

Note that the only thing needed to get this result is an interaction
$H^{FW}_I$ which is block--diagonal to a given order in $1/M$.
Thus a further transformation on the Hamiltonian which preserves the
block--diagonal nature to the same order in $1/M$ will also give the same
correct S--matrix $S_{fi}^{eff-FW}$.
Thus it is clear that the Foldy--Wouthuysen algorithm is not unique.
It has been observed by Barnhill and others (see \cite{Barnhill} and
references therein) that the order by which different, noncommuting
odd operators are eliminated from the relativistic Hamiltonian may lead
to such a ``freedom'' in the choice of the FW Hamiltonian.

In contrast to the Foldy--Wouthuysen reduction, the direct Pauli reduction
leads
to a $4 \times 4$ Hamiltonian $H^P_I=T_0 H_I T_0^\dagger$ which is not
block--diagonal in general. Thus the terms in $S_{fi}$ involving
negative--energy
states will not vanish. Thus if we use $H^{eff-P}_I$ consistently in a
nonrelativistic scattering theory we will get an effective S--matrix
$S_{fi}^{eff-P}$ which is not equal to the (correct) relativistic result
and which is therefore different from that obtained using $H^{eff-FW}_I$, by
virtue
of the $\Phi_{0p}^{FW(-)}$ terms in eq.~(\ref{effsfi2}). In other words,
using $H^{eff-P}_I$ in a nonrelativistic approach amounts to omitting
the negative--frequency contribution of the covariant calculation.
One may expect that this approximation will lead to unacceptable results
in situations where the original interaction Hamiltonian in the
Dirac representation produces a significant
coupling between upper and lower components.

We conclude from this that in cases in which both interactions are known
relativistically, so that one can do a Foldy--Wouthuysen reduction on the
complete interaction, as for example Compton scattering, use of $H^{eff-FW}_I$
gives the correct result to given order in $1/M$, the same result as would be
obtained
in the complete relativistic theory.  In contrast the Pauli reduction effective
Hamiltonian $H^{eff-P}_I$ gives different, and incorrect, results.

In many practical cases however one knows only one of the interactions
relativistically. Thus for example in proton--proton bremsstrahlung or in the
electrodisintegration of nuclei the electromagnetic interaction is
known relativistically but the strong interaction is not. In such
cases the reduction is done on the known interaction alone, and one
cannot generate the contact terms which involve both
interactions. Thus it appears that for these cases one cannot prove
rigorously that one effective Hamiltonian is more correct than
another\footnote{Of course there may be other guiding principles,
such as gauge invariance, which put a constraint on the form
of the effective Hamiltonian.}.

To see how this works in a still somewhat simplified case,
suppose that the interaction part of the Dirac
Hamiltonian is given as before by harmonic interactions of the form
$H_I = H_{1a} + H_{1b}$. Then by carrying out exactly the same steps used to
evaluate
the $\Delta H \times \Delta H$ term in eq.~(\ref{deltas1}) we find for
the part of the full relativistic S--matrix which is first--order in
each of the interactions
\begin{eqnarray}
\label{relsmat}
 S_{fi} & = &
-2 \pi i \delta (E_f + \omega_b - E_i - \omega_a) \int d^3p
\times \nonumber \\ & & \left [ \sum_{\rm spins} \left \{
\frac{<\Phi_{0f} | H_{1b} | \Phi^{(+)}_{0p}>
<\Phi^{(+)}_{0p} | H_{1a} | \Phi_{0i}>}{E_f + \omega_b
-E_p+i\epsilon} \right. \right. \nonumber \\ & & \left. \left. \mbox{} +
\frac{<\Phi_{0f} | H_{1b} | \Phi^{(-)}_{0p}>
<\Phi^{(-)}_{0p} | H_{1a} | \Phi_{0i}>}{E_f + \omega_b
+E_p-i\epsilon} \right \} + \quad \mbox{c.~t.} \right ].
\end{eqnarray}

We suppose that we do not know the interaction $H_{1b}$ in a
relativistic form. This means that we must approximate the full matrix
elements \mbox{$<\Phi_{0f} | H_{1b} | \Phi^{(+)}_{0p}>$} and $<\Phi_{0f} |
H_{1b} | \Phi^{(-)}_{0p}>$ in some way. Typically one would neglect
the latter and evaluate the former by using a nonrelativistic matrix
element calculated, say, from a potential. Then one would reduce the known
interaction $H_{1a}$ to an effective Hamiltonian.

Using eqs.~(\ref{Heqn2}) and (\ref{dh1}) one has $H_{1a}=T_0^\dagger
H_{1a}^P T_0$ in the Pauli reduction case and $H_{1a} =   T_0^\dagger
H^{FW}_{1a} T_0 - \Delta H^{\prime}_{1a}$ with $\Delta H^{\prime}_{1a}
=  i[S_{1a},H_0] - \dot{S}_{1a}$ in the Foldy--Wouthuysen case. Substituting
these into
eq.~(\ref{relsmat}) above and using eq.~(\ref{deltah1}) to evaluate the
$\Delta H$ terms we find for the part in brackets
\begin{eqnarray}
\label{aproxsmat}
\lefteqn { \Bigg \{
\frac{<\Phi_{0f} | H_{1b} | \Phi^{(+)}_{0p}>
<\Phi^{FW(+)}_{0p} | H^{eff-FW}_{1a} | \Phi^{FW}_{0i}>}{E_f +
\omega_b -E_p+i\epsilon} } \nonumber \\
& & \mbox{} -<\Phi_{0f} | H_{1b} | \Phi^{(+)}_{0p}>
<\Phi^{FW(+)}_{0p} | T_0 iS_{1a} T^\dagger_0 | \Phi^{FW}_{0i}>
\nonumber \\ & & \mbox{} -
<\Phi_{0f} | H_{1b} | \Phi^{(-)}_{0p}>
<\Phi^{FW(-)}_{0p} | T_0 iS_{1a} T^\dagger_0 | \Phi^{FW}_{0i}>
\Bigg \}.
\end{eqnarray}
The result using $H^{eff-P}_{1a}$ is of the same form, except that the
second term does not appear and $H^{eff-FW}_{1a}$ is replaced by
$H^{eff-P}_{1a}$.

The first term in this expression is what one would usually use in
second--order perturbation theory in
cases where only one of the interactions is known. It would lead to an
approximate S--matrix,  $S^{approx}_{fi}$, which differs from the full
relativistic S--matrix by contact terms. Using an expansion analogous to
eq.~(\ref{newschroedingerexp}) and explicit forms for the leading parts
of $S_{1a}$ and $T_0$ it is
straightforward to show that the leading term in $T_0 iS_{1a}
T^\dagger_0 $ is odd and of order $\OI /M$ whereas the first even term
is down by one power of $1/M$. Thus the contact
term involving the negative--energy intermediate states is likely to be
the most important, whereas the one involving positive--energy
intermediate states (which appears only for $H^{eff-FW}_{1a}$) is down
by $1/M$. One can see this explicitly in Compton scattering \cite{paper2}.

Above we showed that when both interactions are known $H^{eff-P}_I$
fails to give the correct effective S--matrix because of the neglect of
the terms involving negative--energy intermediate states. On the other hand
$H^{eff-FW}_I$ gives the correct answer, since it is constructed as the upper
left--hand block of a Hamiltonian which does not connect positive-- and
negative--energy
states, and since it contains the contact terms explicitly.
Now when only one of the interactions is known the natural way of
calculating $S^{approx}_{fi}$ in second--order perturbation theory
forces one to neglect these
contributions for {\em both} effective Hamiltonians, and leads to an
error which to leading order is the same in the two cases.

In most practical calculations, particularly when the unknown interaction
is the strong interaction, second--order perturbation theory is not
sufficient.
In such cases, an example being proton--proton bremsstrahlung,
one must sum the strong interaction to all orders.
Thus one normally would replace $<\Phi_{0f}|H_{1b}|\Phi^{(+)}_{0p}>$
in eq.~(\ref{aproxsmat}) by a full T--matrix.
The cross term gives the interaction in the other order and there
is an additional term, called double scattering in proton--proton
bremsstrahlung, which has a T--matrix element on both sides of
the matrix element of $H^{eff}_{1a}$.
An alternative, but essentially equivalent approach which is often
used is simply to take matrix elements of $H^{eff}_{1a}$ between
eigenstates of the free--plus--strong Hamiltonian.

Neither of these approaches will in general give the correct (relativistic)
answer, as long as the effective operator $H^{eff}_{1a}$ is obtained as a
reduction of the relativistic Hamiltonian between {\em free} states, because
of inconsistency of the operator and the wave functions.
In other words, approximations will be involved in getting from the full
relativistic S--matrix to the effective nonrelativistic one.
Regardless of the choice of $H^{eff}_{1a}$, contact terms will be missed.

It appears then that here, in contrast to the situation when both
interactions are known relativistically, we cannot argue rigorously
that $H^{eff-FW}_I$ is any more correct than $H^{eff-P}_I$. At best it
may be preferred simply because it is no worse an approximation when
the interaction is incompletely known and is in principle the correct
choice when the interaction is completely known.

\section{Transformation of the S--Matrix under a Unitary Transformation}
\label{5}
In the previous section we have explicitly shown to second--order perturbation
theory that the S--matrix remains invariant under a unitary transformation of
the states\footnote{In fact we can even allow for more general
transformations (see also ref.~\cite{Kamefuchi}).
It turns out that the following proof is still correct
if we demand $T(t)$ to be an invertible transformation with the property
of eq.~(\ref{tint}).
Note however that $T_0$ remains a time--independent {\em unitary}
transformation.
One then only has to replace $T^{\dagger}$ by $T^{-1}$ in the proof.}.
Here we will show this in general without making use of a perturbative
approach.
We start with the Dirac equation
\begin{equation}
\label{schreq}
i \frac{\partial \Psi(x)}{\partial t} = (H_0 + H_I(x,\xi)) \Psi(x)
= H(x,\xi) \Psi(x),
\end{equation}
where $H_0=\vec{\alpha}\cdot\vec{p}+\beta M$, $x=(x^0,\vec{x})$.
We control the switching on and off of the interaction in the
remote past and distant future through $H_I(x,\xi)=H_I(x) e^{-\xi |t|},
\xi \ge 0$.
The defining equation for the full propagator in the presence of interaction
reads \cite{Bjorken,Itzykson}
\begin{equation}
\label{fullprop}
\left(i\frac{\partial}{\partial t}-H_0-H_I(x,\xi)\right)
S_{F,H_I}(x,y)\beta=\delta^4(x-y),
\end{equation}
and a perturbative solution satisfying the Feynman--St\"uckelberg boundary
condition is given by
\begin{eqnarray}
\label{fullproppert}
\lefteqn{S_{F,H_I}(x,y) =  S_F(x-y) + \int \,d^4x_1 S_F(x-x_1)\beta
H_I(x_1,\xi)
S_F(x_1-y)} \nonumber \\
& & + \int \,d^4x_1 d^4x_2 S_F(x-x_2) \beta H_I(x_2,\xi) S_F(x_2-x_1)
\beta H_I(x_1,\xi) S_F(x_1-y) \nonumber \\
& & + \dots,
\end{eqnarray}
where $S_F(x-y)$ is the free Feynman propagator \cite{Bjorken,Itzykson}.
For $y^0_1<x^0<y^0_2$ the formal solution for $\Psi(x)$  may be constructed as
\begin{equation}
\label{psisol}
\Psi(x) = i\int d^3y_1 \, S_{F,H_I}(x,y_1) \beta \Psi^{(+)}(y_1)
         -i\int d^3y_2 \, S_{F,H_I}(x,y_2) \beta \Psi^{(-)}(y_2),
\end{equation}
where $\Psi^{(+)}(y_1)$ and $\Psi^{(-)}(y_2)$ are superpositions of positive--
and negative--energy solutions of the free Dirac equation, respectively.
If we specify the boundary conditions as
\begin{eqnarray}
\label{bound1}
\lim_{x^0 \rightarrow -\infty} \Psi^{(+)}(x) & = & \Psi^{(+)}_{0i}(x),
\nonumber \\
\lim_{x^0 \rightarrow \infty} \Psi^{(-)}(x) & = & 0,
\end{eqnarray}
where $\Psi^{(+)}_{0i}(x)$ is a positive--energy eigenfunction of the free
Dirac equation with eigenvalue $E_i>0$,
we can either describe the scattering of a particle or pair annihilation
in the potential,
\begin{eqnarray}
\label{pspa}
S_{fi} & = & \lim_{\xi\rightarrow 0_+}\lim_{x^0 \rightarrow \infty}
\int d^3x \, \Psi^{(+)\dagger}_{0f}(x)
\Psi(x) \quad\mbox{for particle scattering,} \nonumber \\
S_{fi} & = & \lim_{\xi\rightarrow 0_+}\lim_{x^0 \rightarrow -\infty}
\int d^3x \, \Psi^{(-)\dagger}_{0f}(x)
\Psi(x) \quad\mbox{for pair annihilation,}
\end{eqnarray}
with $\Psi^{(-)}_{0f}$ a negative--energy solution of the free Dirac
equation.
Likewise, specifying the boundary conditions
\begin{eqnarray}
\label{bound2}
\lim_{x^0 \rightarrow \infty} \Psi^{(-)}(x) & = & \Psi^{(-)}_{0i}(x), \nonumber
\\
\lim_{x^0 \rightarrow -\infty} \Psi^{(+)}(x) & = & 0,
\end{eqnarray}
leads to a description of the scattering of an antiparticle or pair
creation in the potential,
\begin{eqnarray}
\label{apspc}
S_{fi} & = & \lim_{\xi\rightarrow 0_+}\lim_{x^0 \rightarrow -\infty}
\int d^3x \, \Psi^{(-)\dagger}_{0f}(x)
\Psi(x) \quad\mbox{for antiparticle scattering,} \nonumber \\
S_{fi} & = & \lim_{\xi\rightarrow 0_+}\lim_{x^0 \rightarrow \infty}
\int d^3x \, \Psi^{(+)\dagger}_{0f}(x)
\Psi(x) \quad\mbox{for pair creation.}
\end{eqnarray}
When defining the S--matrix elements of eq.~(\ref{pspa}) and (\ref{apspc}),
it is assumed that the system was in an eigenstate of $H_0$ with positive
energy in the remote past or will be in an eigenstate of $H_0$ with
negative energy in the distant future, respectively. It interacts with
the potential at intermediate times and and evolves under the influence
of $H_0$ in the distant future or remote past, respectively.
The S--matrix is the mathematical idealization of extending the interaction
over the complete t--axis \cite{Bogoliubov}. However, it is important
to realize that the limit $\xi \rightarrow 0_+$ has to be taken at the
end (see e.~g.~ref.~\cite{Itzykson}, p 165 f).

If we introduce a unitary transformation $T(x)$ which depends on
the interaction Hamiltonian $H_I(x,\xi)$, or parts of it, with the property
\begin{equation}
\label{tint}
\lim_{H_I(x,\xi) \rightarrow 0} T(x) = T_0,
\end{equation}
where $T_0$ is a time--independent unitary transformation, we will show that
\begin{equation}
\label{equiv}
S'_{fi}=S_{fi},
\end{equation}
provided the boundary conditions and $S'_{fi}$ are defined in an analogous way
as in eq.~(\ref{bound1}) - (\ref{apspc}), e.~g.~for particle scattering
\begin{eqnarray}
\label{smatrixp}
\lim_{x^0 \rightarrow -\infty} \Psi'^{(+)}(x) & = & \Psi'^{(+)}_{0i}(x),
\nonumber \\
\lim_{x^0 \rightarrow \infty} \Psi'^{(-)}(x) & = & 0, \nonumber \\
S'_{fi} & = & \lim_{\xi\rightarrow 0_+}\lim_{x^0 \rightarrow \infty}
\int d^3x \, \Psi'^{(+)\dagger}_{0f}(x)
\Psi'(x).
\end{eqnarray}
In eq.~(\ref{smatrixp}) $\Psi'^{(+)}_{0i}(x) $ and $\Psi'^{(+)}_{0f}(x)$ are
positive--energy
eigenstates of $H'_0=T_0 H_0 T^{\dagger}_0$ and $\Psi'(x)$ satisfies the
Dirac equation
\begin{equation}
\label{newdirac}
i \frac{\partial \Psi'(x)}{\partial t} =
(H'_0 + H'_I(x,\xi)) \Psi'(x) = H'(x,\xi) \Psi'(x),
\end{equation}
with $H'(x,\xi)=T(x)H(x,\xi)T^{\dagger}(x)-iT(x)\partial T^{\dagger}(x)
/\partial t$.
The extension to the other cases is straightforward.
We then find for the S--matrix element $S'_{fi}(\xi)$
\begin{eqnarray}
S'_{fi}(\xi) & = & \lim_{x^0 \rightarrow \infty} \int  d^3x \,
                   \Psi^{(+)\dagger}_{0f}(x) T_0^{\dagger} T(x) \Psi(x).
\end{eqnarray}
For any arbitrarily small but finite $\xi$ eq.~(\ref{tint}) leads to
\begin{equation}
\label{limits}
\lim_{x^0 \rightarrow \infty} T^{\dagger}_0 T(x)=1
\end{equation}
and we therefore obtain
\begin{equation}
\label{equivxi}
S'_{fi}(\xi) = S_{fi}(\xi).
\end{equation}
Taking the limit $\xi \rightarrow 0_+$ in eq.~(\ref{equivxi}) yields the
desired result,
eq.~(\ref{equiv}).

The above derivation is extremely simple, but it does not explicitly reveal
the importance of $H'(x,\xi)$ of eq.~(\ref{newdirac}), in particular,
it does not show why it is $H'(x,\xi)$ and {\em not}
$T(x)H(x,\xi)T^{\dagger}(x)$
which enters the calculation of the primed S--matrix element.
For that reason we provide a slightly more complicated derivation which,
however, gives more insight into the role played by $H'(x,\xi)$.
We once again only discuss the case of particle scattering.
Using eq.~(\ref{psisol}) and the boundary condition of eq.~(\ref{smatrixp})
we may write $S'_{fi}$ as
\begin{eqnarray}
\label{proof}
S'_{fi} & = & i \lim_{\xi\rightarrow 0_+}
\lim_{\stackrel{x^0 \rightarrow \infty}{y^0 \rightarrow -\infty}}
\int d^3x d^3y \, \Psi'^{(+)\dagger}_{0f}(x) S'_{F,H'_I}(x,y) \beta
\Psi'^{(+)}_{0i}(x)  \nonumber \\
& =  & i  \lim_{\xi\rightarrow 0_+}
\lim_{\stackrel{x^0 \rightarrow \infty}{y^0 \rightarrow -\infty}}
\int d^3x d^3y \, \Psi^{(+)\dagger}_{0f}(x) T^{\dagger}_0 S'_{F,H'_I}(x,y)
\beta T_0 \Psi^{(+)}_{0i}(x),
\end{eqnarray}
where $S'_{F,H'_I}(x,y)$ is defined by an equation analogous to
eq.~(\ref{fullprop}), namely
\begin{equation}
\label{fullpropp}
\left(i\frac{\partial}{\partial t}-H'_0-H'_I(x,\xi)\right)
S'_{F,H'_I}(x,y)\beta=\delta^4(x-y).
\end{equation}
Clearly, at this point one realizes that $H'(x)$ rather than
$T(x)H(x,\xi)T^{\dagger}(x)$ is the relevant operator in
the defining equation for the full propagator.
It may easily be shown that
\begin{equation}
\label{sup}
S'_{F,H'_I}(x,y) \beta = T(x) S_{F,H_I}(x,y) \beta T^{\dagger}(y)
\end{equation}
solves the equation of motion for $S'_{F,H'_I}$, eq.~(\ref{fullpropp}),
provided $S_{F,H_I}$ satisfies eq.~(\ref{fullprop}).
We then insert eq.~(\ref{sup}) into eq.~(\ref{proof}) to obtain
\begin{eqnarray}
\label{proofc}
S'_{fi} & =  & i  \lim_{\xi\rightarrow 0_+}
\lim_{\stackrel{x^0 \rightarrow \infty}{y^0 \rightarrow -\infty}}
\int d^3x d^3y \, \Psi^{(+)\dagger}_{0f}(x) T^{\dagger}_0 T(x)
S_{F,H_I}(x,y) \beta T^{\dagger}(y) T_0 \Psi^{(+)}_{0i}(x). \nonumber \\ & &
\end{eqnarray}
Once again we make use of eq.~(\ref{tint}) for any arbitrarily
small but finite $\xi$,
\begin{equation}
\label{ad}
\lim_{y^0 \rightarrow -\infty} T^{\dagger}(y) T_0 = 1 = \lim_{x^0 \rightarrow
\infty} T_0^{\dagger} T(x),
\end{equation}
and obtain
\begin{eqnarray}
\label{proofcc}
S'_{fi}(\xi) & =  & i
\lim_{\stackrel{x^0 \rightarrow \infty}{y^0 \rightarrow -\infty}}
\int d^3x d^3y \, \Psi^{(+)\dagger}_{0f}(x) S_{F,H_I}(x,y) \beta
\Psi^{(+)}_{0i}(x) \nonumber \\
& = & S_{fi}(\xi),
\end{eqnarray}
which is identical with eq.~(\ref{equivxi}).

In conclusion, we have provided two nonperturbative arguments for the
S--matrices to be equal provided the states are related by a unitary
transformation.

\section{Summary and Conclusions}

We have addressed in this paper the question of how one obtains
effective nonrelativistic Hamiltonians from known, or partially known,
relativistic interactions. Such Hamiltonians are important because
they allow the use of nonrelativistic formalisms, and make
connection with interactions which may be known only in some
nonrelativistic approximation, and yet at the same time incorporate
relativistic corrections to some order in $1/M$.

We discussed two different ways of obtaining such effective $2 \times 2$
Hamiltonians.  The first is the Foldy--Wouthuysen interaction which is
obtained by a unitary time--dependent transformation of the states.
This approach decouples the upper from the lower components order by
order in $1/M$ and so, by projecting out the upper left--hand block, leads
naturally to an effective Hamiltonian $H^{eff-FW}_I$ to be
used with two--component positive--energy wave functions.

The second method consists of making a two--component Pauli reduction
of the matrix element of the interaction Hamiltonian in the Dirac
representation between free positive--energy spinors. The result is then
regarded as an effective Hamiltonian $H^{eff-P}_I$ to be used between
nonrelativistic Pauli wave functions. This approach is equivalent to
making a unitary transformation of the states  using the time--independent
transformation $T_0$ and then projecting out the upper
left--hand block to get $H^{eff-P}_I$. Usually the result is expanded in
$1/M$, typically to order $1/M^2$ or $1/M^3$, though in some
applications this step is omitted, i.~e.~the complete expression
without a $1/M$ expansion is used.  Both of these methods for
obtaining effective Hamiltonians and introducing relativistic
corrections have been used for a variety of processes in the
literature.

It has been tacitly assumed in the literature that both approaches
yield the same results, i.~e. that it makes no difference which of the
two effective Hamiltonians is used.  We have seen that this is {\em
in general} not true and that in fact it does make a difference which
is used, at least in the case when the initial relativistic
interaction is known completely.

For first--order matrix elements differences in the results obtained
using the two different effective Hamiltonians arise only when at least one
state is off--energy--shell, or in other words only when energy is not
conserved, as
the difference of the matrix elements calculated with different
effective Hamiltonians is proportional to the sum of initial energies
minus the sum of the final energies at a vertex.

However, in higher--than--first--order, time--ordered perturbation theory,
energy is not conserved at each individual vertex. Furthermore
intermediate negative--energy states contribute and so we might expect
effects from the projection eliminating the effects of the negative--energy
states which is used in getting the effective Hamiltonians from
the relativistic ones.  We found in second--order perturbation theory
that the effective S--matrices, obtained by using the $2 \times 2$
effective Hamiltonians in usual nonrelativistic scattering theory were
in fact different. The result obtained using $H_I^{eff-FW}$ agreed with
the full relativistic S--matrix while that obtained with $H_I^{eff-P}$
did not.

At first glance this seems inconsistent with the perception that the
observables should not change under a unitary transformation. As we
showed, both in second--order perturbation theory and via a general
argument, the full relativistic S--matrix is invariant under such
time--dependent unitary transformations, of which $T(t)$ leading to
$H_I^{FW}$ and $T_0$ leading to $H_I^P$ are examples.

This apparent conflict was resolved via an understanding of the
additional approximations necessary in going from the full
relativistic S--matrix to the effective S--matrix calculated
nonrelativistically. To get $S_{fi}^{eff}$ one must neglect the
negative--energy intermediate states. Since $H_I^{FW}$ is block--diagonal
and does not connect positive-- to negative--energy states anyway,
this approximation has no effect and in this case the effective
S--matrix $S_{fi}^{eff-FW}$ is the same as the full relativistic one. In
contrast $H_I^P$ does connect negative-- and positive--energy states. Thus the
additional approximation needed to get $H_I^{eff-P}$, namely the
projection for positive energies which picks out just the upper left--hand
block, leads to the neglect of some nonzero terms and thus to an
effective S--matrix $S_{fi}^{eff-P}$ which is different from the full
relativistic S--matrix.

The results just described apply only to cases where the full
relativistic interaction is known, as for example Compton scattering
or pion photoproduction.  In many practical situations only part of
the relativistic Hamiltonian is known.  Typically, such a situation
may involve a nonrelativistic potential approach for the strong
interaction and at the same time a relativistic treatment of the
electromagnetic interaction (including the anomalous magnetic moment
and possibly on--shell form factors).

In such cases the situation is ambiguous. Lack of knowledge of one of
the interactions makes it impossible to calculate the contact terms
explicitly and hence they must be neglected whichever effective
Hamiltonian is used. Use of $H_I^{eff-FW}$ requires one to drop an
additional contact term, which is however of order $1/M$ times the
main contact terms which are dropped in all cases. For some processes gauge
invariance allows one to construct non--unique contact terms which can be added
in by hand. However in general there may be
numerical differences when different effective Hamiltonians are used.
Unfortunately it does not seem to be possible to
determine in a rigorous way which is best for these situations, though
there may be a philosophical predilection to use the Foldy--Wouthuysen
approach even in the absence of full information as that is the
correct approach when the interaction is fully known.

In conclusion then, our analysis indicates that when the interaction
is fully known relativistically use of the Foldy--Wouthuysen reduction
in order to incorporate relativistic corrections to a nonrelativistic
treatment leads to correct results. In contrast using the matrix element of
the interaction Hamiltonian in the Dirac representation between free
positive--energy states as an effective nonrelativistic Hamiltonian
gives incorrect results. However when only part of the interaction is
known, as is the case in many practical situations, both methods miss
the leading order contact terms. Thus in this situation one cannot
argue rigorously that one should be preferred over the other.

\section*{Acknowledgements}

This work was supported in part by a grant from the Natural Sciences and
Engineering Research Council of Canada.  The authors would like to thank
J.~H.~Koch for for useful comments.

\newpage

\begin{figure}
\label{figppg}
\end{figure}
{\bf Figure 1:} Cross section and analyzing power for proton--proton
bremsstrahlung for
an incident laboratory energy of 280 MeV and a coplanar equal angle geometry
for the outgoing protons, plotted versus the photon angle. The calculations
were done using the full Bonn potential and the formalism of
ref.~\protect{\cite{Workman}} except that Coulomb corrections are not included.
The solid
(dotted) curve corresponds to using \mbox{$H_I^{eff-FW}$}
\mbox{($H_I^{eff-P}$)} for the electromagnetic interaction.  Note the
suppressed zeros on both axes.

\end{document}